\documentclass[twocolumn,showpacs,preprintnumbers,amsmath,amssymb]{revtex4}
\usepackage{graphicx}
\usepackage{psfig}
\usepackage{dcolumn}
\usepackage{bm}
\usepackage{multirow}
\usepackage{feynmf}
\begin{document}


\title{Correlation energy contribution to nuclear masses}

\author{S.Baroni$^{a}$, M.Armati$^{a}$, F.Barranco$^{c}$, R.A.Broglia$^{a,b,d}$,
        G.Col\`o $^{a,b}$, G.Gori$^{a,b}$ and E.Vigezzi$^{b}$} 
\affiliation{
$^a$ Dipartimento di Fisica, Universit\`a degli Studi di Milano,via Celoria 16, 20133 Milano, Italy.\\
$^b$ INFN, Sezione di Milano, via Celoria 16, 20133 Milano, Italy.\\
$^c$ Departamento de Fisica Aplicada III, Escuela Superior de Ingenieros, camino de los Descubrimientos s/n,  
41029 Sevilla, Spain.\\
$^d$ The Niels Bohr Institute, University of Copenhagen, Blegdamsvej 17, 2100 Copenhagen \O, Denmark.}

\date{\today}




\begin{abstract}
The ground state correlation energies associated with collective surface and pairing vibrations are
calculated for Pb- and Ca-isotopes. It is shown that this contribution, when added to those predicted 
by one of the most accurate modern nuclear mass formula (HFBCS MSk7  mass formula), reduces the associated
rms error by an important factor, making mean field theory, once its time dependence
is taken into account, a quantitative predictive tool for nuclear masses.
\end{abstract}

\maketitle

Many of the nuclei in atoms we find in nature are processed in the stars. In particular, heavy elements far beyond
$^{56}Fe$ are formed through a chain of neutron capture reactions with subsequent beta decay. The source of the rather
low energy neutrons are reactions like ($\alpha$,n) on $^{22}Ni$ and $^{13}C$ where the alphas are highly abundant
during He burning in the stars. The small neutron abundance then causes a particular path to be followed which lies 
mainly along the valley of beta stability as the capture process follows a time scale controlled 
by the variety of $\beta$-decay processes.
The end point of this slow processes (s-process) is above $^{209}Bi$ via $\alpha$-decay. There is
clear evidence for sharp peaks in the abundance curve near neutron closed shell configurations which
commonly result from the existence of slow neutron capture processes.
However, the abundance curve displays a number of additional peaks at around 8-12 mass units prior to the
s-process abundance peaks. This indicates that other capture processes where neutron capture is 
involved have to take place to produce the corresponding elements as well as elements beyond 
$^{209}Bi$. This is called the rapid or r-process. It is known that the r-process path evolves 
along the neutron drip line region \cite{1}, where the binding energy of the least bound neutron
is quite small (few hundred KeV) in the presence of a large neutron flux \cite{Rol.Ro:1988}.
The need for a formula able to predict known masses with an accuracy of this order of magnitude seems quite
natural. On the other there still exist many uncertainties in extrapolating our present-day knowledge
of formulas which predict nuclear masses, towards the region of the drip line
where weakly bound nuclei are found. In fact, the best account of the experimental data
based on mean-field theory provides a fitting to the 2135 measured masses with N,Z $>$ 8
with a rms error of 0.674 MeV \cite{Gor.ea:2002}. 
This has been achieved by means of Hartree-Fock-Bardeen-Cooper-Schrieffer 
(HFBCS) calculations which employ a Skyrme-type zero-range effective force 
in the mean field channel, supplemented by a zero-range pairing interaction.
The 14 parameters set is named BSk2. As a reference point for our work, we have
considered a parameter set of almost equal quality, denoted by MSk7, where
the rms error is 0.738 MeV \cite{Gor.ea:2001}.  
Although this accuracy is remarkable, one is still
not satisfied. In fact, while it takes tens of MeV to remove two nucleons from a nucleus 
lying along the valley of stability, this quantity becomes as low as 300 keV in halo nuclei
like $^{11}Li$ \cite{Li}. The need for a mass formula at least a factor of two 
more accurate than the MSk7 mass formula is evident.

Nuclei display both single-particle and collective degrees of freedom. Consequently,
the corresponding ground states and associated nuclear masses reflect the zero point
fluctuations (ZPF) associated with these modes. While mean field theory includes 
fluctuations associated with quasiparticles, it is only time dependent mean field theory which 
takes into account the zero point fluctuations associated with collective modes.
Despite this, the information in the literature concerning ground state correlation
energies associated with collective modes is rather scarce. A recent
suggestion to consider their effect was put forward  
by G.F. Bertsch and K. Hagino \cite{BH}. Later, realistic calculations for the quadrupole
degree of freedom were
performed for light nuclei \cite{SJ} and for a few selected 
isotopes within the so-called Generator Coordinate Method (GCM) \cite{BBH}.
 
In the present paper we work out, making use of  
Random Phase Approximation (RPA) calculations, the ground state 
correlation energies
associated with both surface (quadrupole and octupole modes) and pairing vibrations
for the Ca- and Pb-isotopes.
Because pairing vibrations \cite{Bes.Br:1966} have a collective character only around closed shell nuclei
(being essentially pure two-quasiparticle states lying on top of twice the pairing gap in 
superfluid systems, cf. also \cite{And:1958}), one expects the associated ZPF to lead to important
corrections to the mass formula \cite{Gor.ea:2002}. This in keeping with the fact that the largest deviation
from experiment found in this mass formula are observed in closed shell systems.
It will be concluded that ZPF are important, in particular those associated with pairing vibrations
in closed shell nuclei, in reducing the rms of the Skyrme-HFBCS mass formulas.

To derive the particle-hole RPA equations use can be made of the quasi-boson approximation
where the particle-hole operators $a^{+}_{k}a_{i}$ and $(a^{+}_{k}a_{i})^{+}$ are 
replaced by the boson operators $\Gamma^{+}_{ki}$ and $\Gamma_{ki}$. Because collective 
vibration can be viewed as correlated particle-hole excitations,
the corresponding boson creation operator can be written as

\begin{equation}\label{eq: 10}
    \Gamma^{+}_{\alpha}(n)=\sum_{ki}(X_{ki}^{\alpha}(n)\Gamma^{+}_{ki}+
                           Y_{ki}^{\alpha}(n)\Gamma_{ki}),
\end{equation}
where $X_{ki}^{\alpha}$ and $Y_{ki}^{\alpha}$ are the forwardsgoing and 
backwardsgoing amplitudes fullfilling the normalization condition

\begin{equation}\label{eq: 15}
    [\Gamma_{\alpha}(n),\Gamma^{+}_{\alpha}(n)]=
      \sum_{ki}({|X_{ki}^{\alpha}|}^{2}-{|Y_{ki}^{\alpha}|}^{2})=1.
\end{equation}
The equations which determine the frequencies of the vibrational modes of quantum number $\alpha$
(with progressively high energy $n=1,2,\ldots$) are obtained from the relation

\begin{equation}\label{eq: 20}
    [H,\Gamma^{+}_{\alpha}(n)]=\hbar\omega_{\alpha}(n)\Gamma^{+}_{\alpha}(n).
\end{equation}

In the above equation $H$ is the total hamiltonian, sum of a single-particle and a
two-body interaction term. In the present calculations we have 
used self-consistently the Skyrme plus pairing interaction given by the 
MSk7 parameter set. The RPA ground state energy is given by (cf. e.g. \cite{Rin.Sc:1980})

\begin{equation}\label{eq: 30}
    E_{RPA}=E_{HF}-(2\lambda + 1)\sum_{\alpha,n}\hbar\omega_{\alpha}(n)
                   \sum_{ki}\left| Y_{ki}^{\alpha}(n) \right|^{2},
\end{equation}
in keeping with the fact that the amplitudes $Y_{ki}^{\alpha}(n)$ are directly
related to the ground state correlations induced by the corresponding vibrational modes.
The second term of the r.h.s. is called correlation energy. 

It is well known that open shell nuclei display a finite BCS gap $\Delta$, as
a result of the pairing correlations acting between pairs of nucleons moving
in time reversal states close to the Fermi energy. In closed shell nuclei, pairing
correlations are not strong enough to overcome the large single-particle gap, and
$\Delta=0$. Nevertheless, pairing plays an important role in determining the 
(low-lying) structure of these nuclei. In particular, it leads to pairing 
vibrations \cite{Bes.Br:1966}, which change in two the number of nucleons,
that is vibrations which can be viewed as two correlated particles (pair addition modes) 
or two correlated holes (pair removal modes).
An example of these modes is provided by the ground state of $^{210}Pb$ and of $^{206}Pb$ which can be viewed as the
pair addition and the pair subtraction modes of $^{208}Pb$ respectively. These modes
are specifically probed through two particle transfer reactions \cite{Bro.ea:ADV}. 
The pair addition and pair subtraction modes can be written as

\begin{equation}\label{eq: 40}
    \Gamma^{+}_{a}(n)=\sum_{k}X_{k}^{a}(n)\Gamma^{+}_{k}+
                   \sum_{i}Y_{i}^{a}(n)\Gamma_{i},
\end{equation}
and
\begin{equation}\label{eq: 50}
    \Gamma^{+}_{r}(n)=\sum_{i}X_{i}^{r}(n)\Gamma^{+}_{i}+
                   \sum_{k}Y_{k}^{r}(n)\Gamma_{k},
\end{equation}
where $\Gamma^{+}_{k}=[a^{+}_{k}a^{+}_{k}]_{0}$ creates a pair of nucleons coupled to angular momentum zero in levels
with energy larger than the Fermi energy ($\varepsilon_{k}>\varepsilon_{F}$), while
$\Gamma_{i}=[a_{i}a_{i}]_{0}$ annihilates a pair of nucleons in occupied states
($\varepsilon_{i}\leq \varepsilon_{F}$).
Use is made of a hamiltonian
   $ H=H_{sp}+H_{p}$,
sum of a single-particle term and of a pairing force with constant matrix
   $ H_{p}=-G\sum_{jj'}a^{+}_{j^{\prime}}a^{+}_{\tilde{j^{\prime}}}a_{\tilde{j}}a_{j}$,
the commutation relation given in Eq.(\ref{eq: 20}) leads to the dispersion relations
\begin{equation}\label{eq: 80}
    \frac{1}{G}=\sum_{k}\frac{\Omega_{k}}{2e_{k}-\hbar\omega_{a}(n)}+
                 \sum_{i}\frac{\Omega_{i}}{2e_{i}+\hbar\omega_{a}(n)},
\end{equation}
\begin{equation}\label{eq: 90}
    \frac{1}{G}=\sum_{k}\frac{\Omega_{k}}{2e_{k}+\hbar\omega_{r}(n)}+
                 \sum_{i}\frac{\Omega_{i}}{2e_{i}-\hbar\omega_{r}(n)},
\end{equation}
where $\Omega=(2j+1)/2$ is the pair degeneracy of the single-particle orbital with
total angular momentum $j$, while $e_{j}=\varepsilon_{j}-\varepsilon_{F}$.
The amplitudes are
\begin{equation}\label{eq: 100}
    X_{k}^{a}(n)=\frac{\Lambda_{a}(n)\sqrt{\Omega_{k}}}{2e_{k}-\hbar\omega_{a}(n)},\quad
    Y_{i}^{a}(n)=\frac{\Lambda_{a}(n)\sqrt{\Omega_{i}}}{2e_{i}+\hbar\omega_{a}(n)},
\end{equation}
and
\begin{equation}\label{eq: 110}
    X_{i}^{r}(n)=\frac{\Lambda_{r}(n)\sqrt{\Omega_{i}}}{2e_{i}-\hbar\omega_{r}(n)},\quad
    Y_{k}^{r}(n)=\frac{\Lambda_{r}(n)\sqrt{\Omega_{k}}}{2e_{k}+\hbar\omega_{r}(n)},
\end{equation}
$\Lambda_{a}(n)$ and $\Lambda_{r}(n)$ being normalization constants determined from the relation 
given in Eq. (\ref{eq: 15}).

\begin{figure}[!h]
\includegraphics[width=0.2\textwidth]{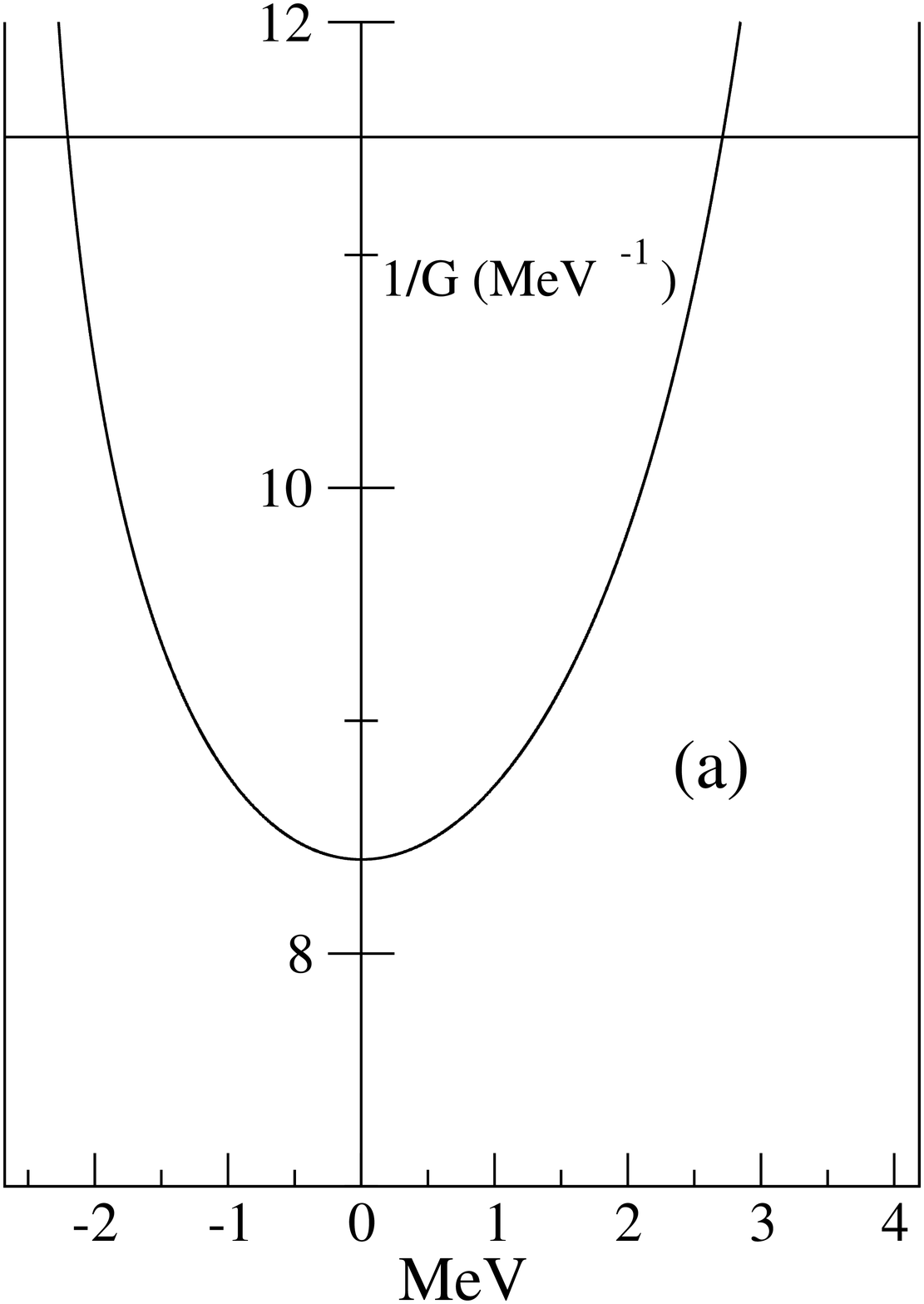}\quad\quad
\includegraphics[width=0.2\textwidth]{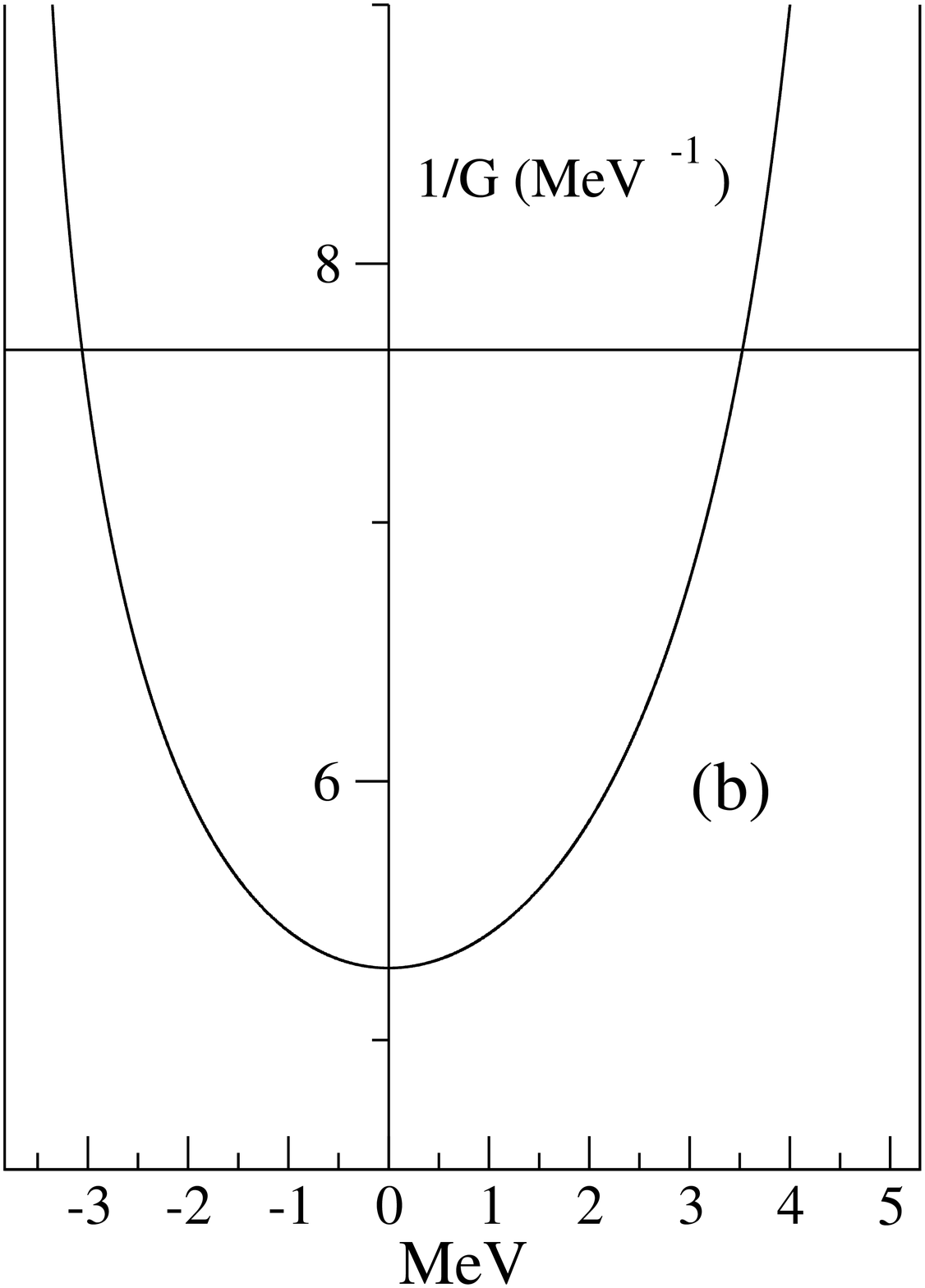}
\caption{Monopole-pairing-vibrations dispersion relation for (a) neutrons and (b) protons 
for the nucleus $^{208}Pb$}
\label{fig: 1}
\end{figure}

In Fig. \ref{fig: 1} we show the dispersion relations given in  Eqs.(\ref{eq: 80}) and
(\ref{eq: 90}) calculated for $^{208}Pb$ for both protons and neutrons (cf. also \cite{Bes.Br:1966}), 
making use of the valence orbitals of this nucleus.
The valence orbitals used in our calculations were determined with the help
of a Woods-Saxon potential with standard parametrization \cite{Boh.Mo:1969} and the energies
have been replaced with the experimental values whenever available. The results obtained using the standard 
Woods-Saxon levels coincide with those calculated making use of the experimental energies within 2\%. 
Making use of the fact that the sum of the pairing binding energies of $^{206}Pb$
and $^{210}Pb$ as well as in $^{206}Hg$ and $^{210}Po$ are $\approx$ 2 MeV (in this
last case one has to take into account the Coulomb repulsion between the two protons, cf. e.g. \cite{Bor.ea:1977}),
one obtains the values of 2.7 MeV and 2.2 MeV for the neutron pair addition and pair 
removal energies, the corresponding values for the proton channel being
3.5 MeV and 3.1 MeV respectively (cf. refs \cite{Bes.Br:1966}, \cite{Bor.ea:1977} and \cite{Bes.Br:1971}).
The associated $Y$'s amplitudes are displayed in Tables \ref{table: 1} and \ref{table: 2}.
\renewcommand{\multirowsetup}{\centering}
\begin{table}[!h]
\begin{tabular}{|c|c|r|r||c|r|}
   \cline{2-6}
    \multicolumn{1}{c|}{} & \multicolumn{3}{c||}{$\lambda^{\pi}=0^{+}$} & \multicolumn{2}{c|}{$\lambda^{\pi}=2^{+}$} \\ \hline 
   \multirow{8}{4mm}{k}
       & \multicolumn{1}{c}{} & \multicolumn{1}{c}{$Y^{r}_{k}$} & \multicolumn{1}{c||}{$X^{a}_{k}$} &
               \multicolumn{1}{c}{}  & \multicolumn{1}{c|}{$X^{a}_{k}$} \\ \cline{2-4}
       & $3d_{3/2}$ & 0.0618 & 0.1053 &  \multicolumn{2}{c|}{} \\ \cline{2-6}
       & $2g_{7/2}$ & 0.0882 & 0.1512 &  $2g_{7/2}\otimes 1i_{11/2}$ & 0.1837 \\ \cline{2-6}
       & $4s_{1/2}$ & 0.0480 & 0.0886 &  $1j_{15/2}\otimes 1j_{15/2}$ & -0.1530 \\ \cline{2-6}
       & $3d_{5/2}$ & 0.0915 & 0.1864 & $2d_{5/2}\otimes 2g_{9/2}$ & 0.3005  \\ \cline{2-6}
       & $1j_{15/2}$ & 0.1542 & 0.3259 & $2i_{11/2}\otimes 2i_{11/2}$ & -0.2169 \\ \cline{2-6}
       & $1i_{11/2}$ & 0.1556 & 0.4122 & $2i_{11/2}\otimes 2g_{9/2}$ & -0.1094  \\ \cline{2-6}
       & $2g_{9/2}$ & 0.1774 & 0.8439 &  $2g_{9/2}\otimes 2g_{9/2}$ & -0.8783   \\ \hline \hline

   \multirow{7}{4mm}{i}
       & \multicolumn{1}{c}{} & \multicolumn{1}{c|}{$X^{r}_{i}$} & \multicolumn{1}{c||}{$Y^{a}_{i}$} & 
                \multicolumn{1}{c}{} & \multicolumn{1}{c|}{$X^{r}_{i}$} \\ \cline{2-6}
       & $3p_{1/2}$ & 0.7853 & 0.0839 & $3p_{1/2}\otimes 2f_{5/2}$ & 0.8431 \\ \cline{2-6}
       & $2f_{5/2}$ & 0.4841 & 0.1209 & $3p_{1/2}\otimes 3p_{3/2}$ & 0.3756  \\ \cline{2-6}
       & $3p_{3/2}$ & 0.2879 & 0.0899 & $2f_{5/2}\otimes 2f_{5/2}$ & -0.2614  \\ \cline{2-6}
       & $1i_{13/2}$ & 0.3347 & 0.1402 & $2f_{5/2}\otimes 3p_{3/2}$ & -0.1378  \\ \cline{2-6}
       & $2f_{7/2}$ & 0.1856 & 0.0914 & $3p_{3/2}\otimes 3p_{3/2}$ & -0.1186  \\ \cline{2-6}
       & $1h_{9/2}$ & 0.1461 & 0.0839 & $3p_{3/2}\otimes 2f_{7/2}$ & 0.1422  \\ \hline

\end{tabular}
\caption{RPA wavefunctions of the neutron pair addition (a) and pair
removal (r) modes of $^{208}Pb$ with multipolarities and parity  
$\lambda^{\pi}=0^{+},2^{+}$.}
\label{table: 1}
\end{table}

Inserting these results in Eq.(\ref{eq: 30}), one obtains the
ground state correlation energy values -0.399 MeV (neutrons) and -0.449 MeV (protons) respectively.
In all calculations we have kept the contribution of only the lowest
($n = 1$) pair addition and pair subtraction modes, in keeping with the fact that,
as a rule, the $n\neq 1$ modes are much less collective.
Pairing vibrations with multipolarity $\lambda \neq 0$, in particular quadrupole and hexadecapole
pairing vibrations, have also been identified around closed shell nuclei (cf. \cite{Bro.ea:ADV}, \cite{Bor.ea:1977},
\cite{Bes.Br:1971} and refs therein).
In Table \ref{table: 3} we display the contributions to the ground state energy
(i.e. $E_{RPA}$ as defined in Eq.(\ref{eq: 30})) associated with the monopole,
quadrupole and hexadecapole pair addition and pair removal modes for both
neutrons and protons associated with $^{208}Pb$, the summed contribution amounting to
-1.981 MeV ($\approx -1.196$ MeV $-0.785$ MeV).


\renewcommand{\multirowsetup}{\centering}
\begin{table}[!h]
\begin{tabular}{|c|c|r|r||c|r|}
   \cline{2-6}
    \multicolumn{1}{c|}{} & \multicolumn{3}{c||}{$\lambda^{\pi}=0^{+}$} & \multicolumn{2}{c|}{$\lambda^{\pi}=2^{+}$} \\ \hline 
   \multirow{8}{4mm}{k}
       & \multicolumn{1}{c}{} & \multicolumn{1}{c|}{$Y^{r}_{k}$} & \multicolumn{1}{c||}{$X^{a}_{k}$} & 
                \multicolumn{1}{c}{} & \multicolumn{1}{c|}{$X^{a}_{k}$} \\ \cline{2-4}
       & $3p_{1/2}$ & 0.0431 & 0.0678 & \multicolumn{2}{c|}{} \\ \cline{2-6}
       & $3p_{3/2}$ & 0.0655 & 0.1090 & $1i_{13/2}\otimes 1i_{13/2}$& -0.1097 \\ \cline{2-6}
       & $2f_{5/2}$ & 0.0835 & 0.1439 &  $2f_{5/2}\otimes 1h_{9/2}$  & 0.1587 \\ \cline{2-6}
       & $1i_{13/2}$ & 0.1547 & 0.3331 & $2f_{7/2}\otimes 2f_{7/2}$ & -0.1359  \\ \cline{2-6}
       & $2f_{7/2}$ & 0.1337 & 0.3601 & $2f_{7/2}\otimes 1h_{9/2}$ & 0.1000 \\ \cline{2-6}
       & $1h_{9/2}$ & 0.1822 & 0.8798 &  $1h_{9/2}\otimes 1h_{9/2}$ & -0.9602   \\ \hline \hline

   \multirow{7}{4mm}{i}
       & \multicolumn{1}{c}{} & \multicolumn{1}{c|}{$X^{r}_{i}$} & \multicolumn{1}{c||}{$Y^{a}_{i}$} & 
                \multicolumn{1}{c}{} & \multicolumn{1}{c|}{$X^{r}_{i}$} \\ \cline{2-6}
       & $3s_{1/2}$ & 0.6862 & 0.0780 & $3s_{1/2}\otimes 2d_{3/2}$ & 0.9271 \\ \cline{2-6}
       & $2d_{3/2}$ & 0.5634 & 0.1010 & $2d_{3/2}\otimes 2d_{3/2}$ & -0.2738  \\ \cline{2-6}
       & $1h_{11/2}$ & 0.4463 & 0.1414 & $3s_{1/2}\otimes 2d_{5/2}$ & 0.1814  \\ \cline{2-6}
       & $2d_{5/2}$ & 0.2672 & 0.0939 & $1h_{11/2}\otimes 1h_{11/2}$ & -0.1222  \\ \cline{2-6}
       & $1g_{7/2}$ & 0.1711 & 0.0824 & $2d_{3/2}\otimes 1g_{7/2}$ & 0.1022 \\ \hline

\end{tabular}
\caption{RPA wavefunctions of the proton pair addition (a) and pair
removal (r) modes of $^{208}Pb$ with multipolarities and parity  
$\lambda^{\pi}=0^{+},2^{+}$.}
\label{table: 2}
\end{table}

\begin{table}[!h]
\begin{tabular}{|c|c|c|c|c|c|}
   \hline
   \multicolumn{2}{|c|}{$0^{+}$} & \multicolumn{2}{c}{$2^{+}$} & \multicolumn{2}{|c|}{$4^{+}$} \\ \hline 
   n & p & n & p & n & p \\ \hline
   -0.399 & -0.449 & -0.609 & -0.244 & -0.189 & -0.092 \\ \hline
\end{tabular}
\caption{Ground state correlation energies, arizing from the neutron (n) and protons (p)
monopole, quadrupole and hexadecapole pairing vibrations in $^{208}Pb$.}
\label{table: 3}
\end{table}

\begin{table}[!h]
\begin{tabular}{|c|c|c|c|c|c|}
   \hline
    & $^{204}Pb$ & $^{206}Pb$ & $^{208}Pb$ & $^{210}Pb$ & $^{212}Pb$  \\ \hline
   \emph{p-h vibrations} & -2.793 & -2.709 & -2.237 & -2.801 & -3.173 \\ \hline
   \emph{pairing vibrations} & -0.785 & -0.785 & -1.981 & -0.785 & -0.785 \\ \hline
\end{tabular}
\caption{Ground state correlation energies for the Pb-isotopes.}
\label{table: 4}
\end{table}

\begin{table}[!h]
\begin{tabular}{|c|c|c|c|c|c|}
   \hline
    & $^{40}Ca$ & $^{42}Ca$ & $^{44}Ca$ & $^{46}Ca$ & $^{48}Ca$  \\ \hline
   \emph{p-h vibrations} & -0.886 & -1.418 & -1.606 & -1.391 & -0.547 \\ \hline
   \emph{pairing vibrations} & -4.761 & -2.978 & -3.239 & -3.500 & -5.823 \\ \hline
\end{tabular}
\caption{Ground state correlation energies for the Ca-isotopes.}
\label{table: 5}
\end{table}

\begin{table}[!h]
\begin{tabular}{|c|c|c|}
   \hline
         & \multicolumn{2}{c|}{$\sigma$ (MeV)} \\ \hline
         & ref. \cite{Gor.ea:2001} & ref. \cite{Gor.ea:2001} + RPA \\ \hline
    Pb & 0.646 & 0.543 \\ 
    Ca & 1.200 & 0.466 \\ \hline
    $\bar{\sigma}$ & 0.964 & 0.505 \\ \hline
\end{tabular}
\caption{Root mean square error associate with the HFBCS MSk7 mass formula of
ref. \cite{Gor.ea:2001} and to this formula (with sligthly adjusted parameters) plus 
the correlation contributions associated with surface and pairing vibrations calculated
in the RPA. The quantity $\bar{\sigma}=\left(\frac{\sigma_{Ca}^2+\sigma_{Pb}^2}{2}\right)^{1/2}$
is shown in the last line.}
\label{table: 6}
\end{table}

In Table \ref{table: 4} we collect the corresponding contribution for a number of Pb-isotopes. Because
pairing vibrations are collective modes only around closed shell nuclei, where particles
and holes can be clearly distinguished, becoming non-collective two-quasiparticle modes outside
closed shells (cf. e.g. ref. \cite{Li}) we have considered the contribution of
neutron pairing vibrations only for the closed shell system (while the proton pairing
vibration were taken into account for all isotopes).
Also shown in table \ref{table: 4} are the contribution to $E_{RPA}$ arising from the low-lying
collective particle-hole vibrations calculated making use of the MSk7 interaction
to determine the single particle states and the particle-hole correlated modes
(cf. Eq.(\ref{eq: 10})). Quadrupole and octupole vibrations with energy $<$ 7 MeV, and
exausting $\geq 2 \%$ of the non-energy weighted sum rule were included in the
calculation of $E_{RPA}$. These conditions essentially select the lowest (one-two) states displaying
correlated wavefunctions.

\begin{figure}[!h]
\centerline{\psfig{file=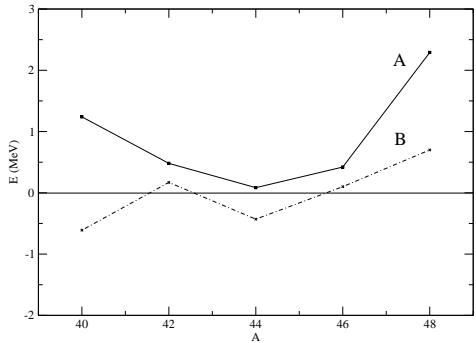,height=4.5cm,angle=0}}
\caption{Difference between the calculated and the experimental ground state energies. The curve (A) 
refers to the HFBCS MSk7 results of ref. \cite{Gor.ea:2001}. The curve (B) was obtained
adding \emph{p-h} and pairing correlations to the curve (A) and renormalizing the Skyrme parameters
as indicated in the text.}
\label{fig: 2}
\end{figure}
Similar calculations were repeated for the calcium isotopes $^{40-48}Ca$.
In this case, the contribution of the proton pairing vibrations calculated for the closed shell systems
$^{40}Ca$ and $^{48}Ca$ were linearly interpolated for the other isotopes, and no
hexadecapole modes were considered. In Table \ref{table: 5} we show the corresponding results, together
with the contribution of the particle-hole vibrational modes.
When adding the results of Tables \ref{table: 4} and \ref{table: 5} to the HFBCS MSk7 mass formula of
Ref. \cite{Gor.ea:2001}, the parameters of the Skyrme interaction should be refitted in order to
provide the best reproduction of experimental masses. This should be done on a large sample of isotopes and
it is beyond the purpose of the present paper. If we restrict ourselves to Ca-isotopes (Pb-isotopes), a slight
renormalization by a factor 0.9964 (0.99945) of the Skyrme parameters $t_{i}$,$W_{0}$ is enough
to shift the calculated value of the masses upwards by $\approx$ 4.3 MeV (3.5 MeV) and minimize the rms deviation, giving
the results displayed in Table \ref{table: 6}, and, for the Calcium isotopes, also shown in Fig. \ref{fig: 2}.
Averaging the rms deviations associated with Ca- and Pb-isotopes, leads
to a value of 0.505 MeV as compared to the value of 0.964 MeV obtained making use of the results of ref. \cite{Gor.ea:2002}.
Although a global readjustment of the mean field parameters should be envisaged, the fact that the locally extracted 
rms deviations have been reduced by a factor of $\approx$ 2 can be considered meaningful.

Similarly to what was observed in assessing the role played by zero point fluctuations 
associated with surface and pairing vibrations in connection with the nuclear mean
square radius \cite{14, 15, 16}, the alignment of rotational nuclei \cite{17} and the pairing phase 
transition as a function of angular momentum \cite{Bro.ea:1968, 18, 19}, we conclude that zero point fluctuations play an important role
in providing the A-dependent contributions to nuclear masses needed to make mean
field theory, including its time dependence, a quantitative predictive tool.



\bibliographystyle{apsrev}

\begin{thebibliography}{99}
\bibitem{1} When neutrons are progressively added to a normal nucleus, the Pauli principle
forces them into states of higher momentum. When the core becomes neutron saturated, 
the nucleus expels most of the wavefunction of the last neutrons outside, to region of
lower density, which because of its larger size can have lower momentum. In most cases neutrons
essentially drip off from the nucleus, defining the lines of stability for neutron number 
in the chart of nuclides (i.e. the neutron drip line for which the neutron binding energy is equal to zero).
\end{thebibliography}

\end{document}